\documentclass[acmsmall]{acmart}
\AtBeginDocument{%
  \providecommand\BibTeX{{%
    \normalfont B\kern-0.5em{\scshape i\kern-0.25em b}\kern-0.8em\TeX}}}

\setcopyright{acmcopyright}
\copyrightyear{2024}
\acmYear{2024}
\acmDOI{10.1145/xxxxxxx.xxxxxxx}

\acmISBN{978-1-4503-XXXX-X/18/06}

\usepackage{multirow}
\usepackage{colortbl}
\usepackage{rotating}
\usepackage{framed}
\usepackage{listings}
\begin{document}

\title{RAG-Enhanced Commit Message Generation}


\author{Linghao Zhang}
\affiliation{%
  \institution{School of Computer Science, Wuhan University}
  \city{Wuhan}
  \country{China}}
\email{starryzhang@whu.edu.cn}

\author{Hongyi Zhang}
\affiliation{%
  \institution{School of Computer Science, Wuhan University}
  \city{Wuhan}
  \country{China}}
\email{harryzhang@whu.edu.cn}

\author{Chong Wang}
\affiliation{%
  \institution{School of Computer Science, Wuhan University}
  \city{Wuhan}
  \country{China}}
\email{cwang@whu.edu.cn}

\author{Peng Liang}
\affiliation{%
  \institution{School of Computer Science, Wuhan University}
  \city{Wuhan}
  \country{China}}
\email{liangp@whu.edu.cn}
\renewcommand{\shortauthors}{Zhang, et al.}


\begin{abstract}
  Commit message is one of the most important textual information in software development and maintenance. However, it is time-consuming to write commit messages manually. Commit Message Generation (CMG) has become a research hotspot. Recently, several pre-trained language models (PLMs) and large language models (LLMs) with code capabilities have been introduced, demonstrating impressive performance on code-related tasks. Meanwhile, prior studies have explored the utilization of retrieval techniques for CMG, but it is still unclear what effects would emerge from combining advanced retrieval techniques with various generation models. This paper proposed \textbf{REACT}, a \textbf{RE}trieval-\textbf{A}ugmented framework for \textbf{C}ommi\textbf{T} message generation. It integrates advanced retrieval techniques with different PLMs and LLMs, to enhance the performance of these models on the CMG task. Specifically, a hybrid retriever is designed and used to retrieve the most relevant code diff and commit message pair as an exemplar. Then, the retrieved pair is utilized to guide and enhance the CMG task by PLMs and LLMs through fine-tuning and in-context learning. The experimental results show that REACT significantly enhances these models’ performance on the CMG task, improving the BLEU score of CodeT5 by up to 55\%, boosting Llama 3’s BLEU score by 102\%, and substantially surpassing all baselines. 
  
\end{abstract}  





\maketitle

\section{Introduction}
In software development and maintenance, the Git version control system has been widely used to store and share code. Within Git, commit messages describe and document the code changes made in a commit. These messages record alterations to the source code, help developers understand the changes in the code, and promote efficient collaboration. Therefore, the commit message serves as one of the critical pieces of textual information in the software engineering life-cycle. However, writing commit messages is time-consuming and laborious for developers. Many developers find them tedious and are not motivated to write~\cite{maa10can}. Additionally, due to the subjectivity of developers, the quality of commit messages in existing code repositories is inconsistent. Some messages can be non-informative (e.g., ``initial commit'', ``last commit today'') or even empty~\cite{changescribe}. As a recent work reported~\cite{whatmakes}, on average 44\% of commit messages did not reach the desired quality, indicating a lack of essential information and the struggle to convey critical details about what the commit did and why. 

In this context, Commit Message Generation (CMG) has become a hot topic in automated software engineering and has garnered attention from this research community. CMG task aims to take the differences between two versions of code as input, typically in the form of a code diff file \texttt{.diff} generated by Git, aka code change, and then generates the corresponding commit message.
Initially, rule-based approaches were used in CMG~\cite{autosumcc,changescribe,deltadoc}, where predefined rules or templates were utilized for generation. Some retrieval-based approaches leverage information retrieval (IR) techniques to suggest commit messages from similar code diff~\cite{nngen, cc2vec}.
With the rapid development of deep learning, many learning-based methods have recently emerged~\cite{cmtgen,nmt,codisum,fira,ptrgncmsg}. These methods treat CMG as a neural machine translation task, in which code diff is the input of neural network model to generate commit messages as output. 
In addition, there are some hybrid approaches~\cite{ATOM,corec,race} that leverage both IR and deep learning techniques. 

More recently, various language models possessing code capabilities have been proposed, including code-specific pre-trained language models such as CodeT5~\cite{codet5} and general large language models, notably ChatGPT\footnote{\url{https://chatgpt.com/}}, which have attracted the vast attention of the research community. In this paper, we refer to these code-capable language models as Code Language Models (CLMs). Different from traditional models, CLMs can be adapted for downstream code-related tasks by merely fine-tuning or prompting~\cite{instruct}, rather than training a model from initial weights. Such approaches not only conserve resources and training time but also can achieve better performance. Researchers have explored applying them to the CMG task~\cite{zhang2024using,cmgreview} and got some promising prospects. 

\begin{figure}[t]
    \centering
    \includegraphics[width=\linewidth]{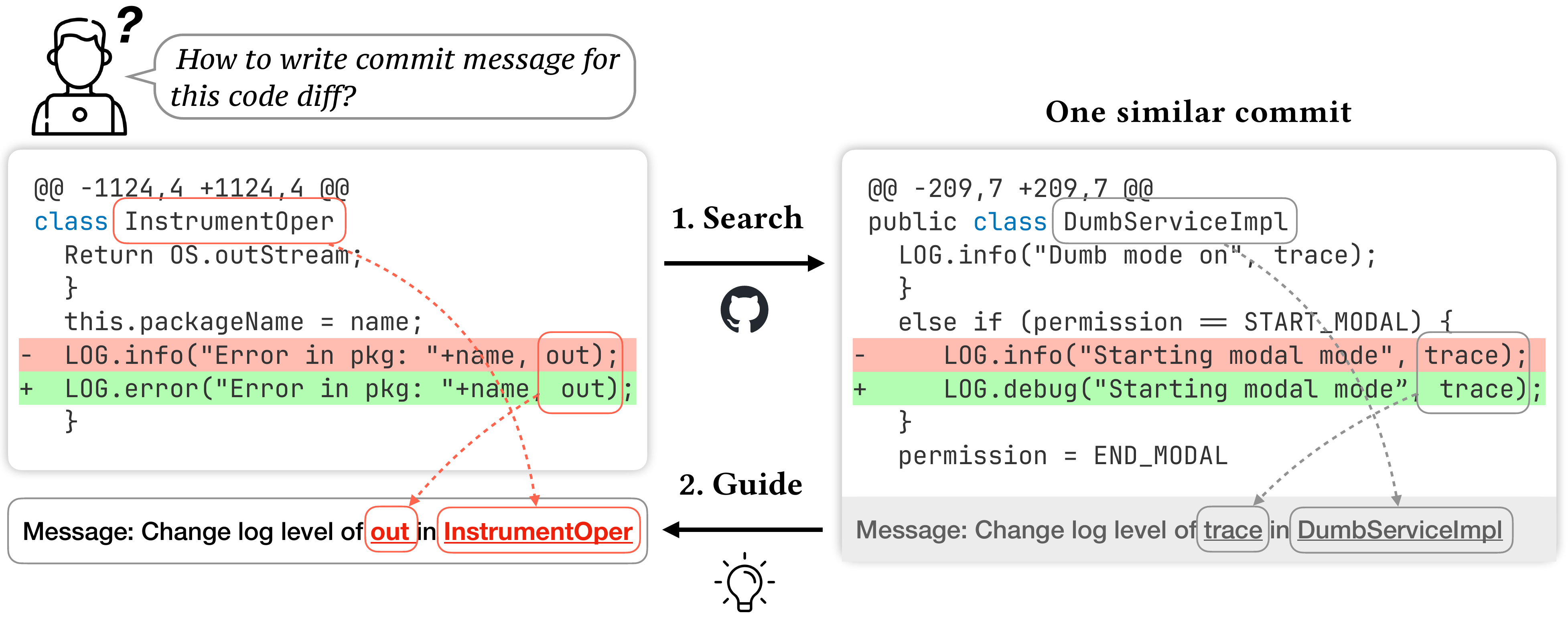}
    \caption{A motivating scenario of how developers write a commit message by referring to a similar exemplar.}
    \label{fig:moti}
\end{figure}

\paragraph{\textbf{Motivation.}} Consider a scenario in Figure~\ref{fig:moti}, where if a developer wants to write a good commit message, one of the efficient ways is to search and imitate the exemplar. That is, exemplary commit messages for similar code diff may be valuable and provide additional information to improve the performance of commit message generation. In essence, this scenario embodies the concept of one paradigm of Retrieval-Augmented Generation (RAG)~\cite{rag}. Prior studies~\cite{race,corec} applied retrieval-augmented approaches to the CMG task and conducted preliminary investigations on the effectiveness of retrieved exemplars in improving generation. However, they opted to train a model from scratch instead of leveraging the prior knowledge embedded in pre-trained CLMs and only focused on specific models and trivial retrieval techniques which limited the performance. Moreover, it still remains unclear how to effectively integrate retrieval techniques with different models and how much improvement such integration could bring.

\paragraph{\textbf{Our approach.}} In this paper, we propose \textbf{REACT} as a \textbf{RE}trieval-\textbf{A}ugmented framework for \textbf{C}ommi\textbf{T} message generation. REACT comprises three phases, i.e., \textit{Retrieve}, \textit{Augment}, and \textit{Generate}, to generate commit messages for given code diff.
\textbf{Retrieve:} We design a hybrid retriever to retrieve the most relevant diff and its corresponding commit message from a comprehensive and high-quality source database. The retrieved diff-message pair will be used to guide the subsequent generation of commit message.
\textbf{Augment:} The query diff and retrieved diff-message pair need to be combined as the input of the models, that is, input augmentation. The augmenter concatenates them with pre-defined special tokens or fills them into a prompt template before passing the result to the generator for the next phase.
\textbf{Generate:} The generator receives augmented inputs and generates a commit message corresponding to the query diff under the guidance of the retrieved pair. We use various CLMs as the backbone of the generator instead of training a model from scratch. This can fully leverage the prior knowledge in CLMs and generate messages more effectively. Some generators with PLM backbones require training to learn how to leverage the additional information provided by the retrieved pair to enhance commit message generation. If LLM is used as the generator, the enhancement can be achieved through in-context learning~\cite{incontext} by embedding the retrieved pairs into the input prompt, eliminating the need for training. The final commit message generated through three stages is enhanced by the additional information or similar patterns provided by the exemplar.
\paragraph{\textbf{Results.}} We designed and conducted comprehensive experiments to evaluate REACT on a widely-used CMG dataset. Before assessing the enhancement effects of REACT, we first reported the performance of directly using CLMs to generate commit messages. The experimental results showed that without integrating the REACT framework, directly generating commit messages with PLMs yielded remarkable results, surpassing all baselines comprehensively. Specifically, CodeT5's BLEU score was 21\% higher than the best baseline method, demonstrating the superiority of PLMs. Although LLMs showed mediocre scores on some metrics, Llama 3's METEOR score still surpassed all baselines without additional training. When integrating CLMs into REACT, the model's performance had broad and significant improvements. CodeT5's BLEU score increased by 55\% compared to direct application, and Llama 3's BLEU score increased by 102\%. The results established the effectiveness and wide applicability of the REACT framework, setting a new SOTA for the CMG task. Furthermore, we conducted a within-project case study on Electron, a popular open-source project from GitHub, where the exemplars were retrieved solely from the project’s historical commits. The results indicated that guided by exemplars, the generation process adheres well to the project’s commit message writing conventions, thereby significantly improving metric scores. This finding provides valuable insights into the application of REACT in real-world within-project CMG scenarios.

\paragraph{\textbf{Contributions.}} The main contributions of our paper can be summarized as follows:

\begin{itemize}
    \item \textbf{An effective retrieval-augmented framework for CMG.} We propose REACT, a retrieval-augmented framework for enhancing commit message generation. REACT can effectively integrate advanced retrieval techniques with various models for better CMG.
    \item \textbf{Extensive experiments are designed to evaluate the performance of various CLMs on the CMG task.} To the best of our knowledge, this work is the first to comprehensively evaluate the performance of various CLMs on the CMG task, including PLMs (e.g., CodeT5, UniXCoder, etc.) and LLMs (e.g., ChatGPT, Llama 3, Gemma). Experimental results show that, benefiting from the code understanding ability acquired during pre-training, CLMs achieved impressive performance on CMG with simple fine-tuning or prompting with basic instruction, surpassing the baselines by a large margin. Specifically, CodeT5's BLEU score was 21\% higher than the best baseline method RACE.
    \item \textbf{Extensive experiments are designed to evaluate the effectiveness of our REACT framework.} We conduct a comprehensive evaluation of REACT, and the results show that REACT can generally enhance the performance of CLMs compared to directly using them for generation. In terms of BLEU scores, REACT improves CodeT5 by up to 55\% and Llama 3 by up to 102\%, respectively.
\end{itemize}

The remainder of this paper is structured as follows: Section~\ref{sec:2} covers related work on CMG and CLMs. Section~\ref{sec:3} introduces our REACT framework, detailing its three phases. Section~\ref{sec:4} describes the experimental setup, including the dataset and model selection, while Section~\ref{sec:5} presents results and analysis. Finally, Section~\ref{sec:6} and Section~\ref{sec:7} discuss the implications and threats to validity respectively, followed by the conclusions in Section~\ref{sec:8}.
The replication package is available online~\cite{data}

\section{Related Work}
\label{sec:2}
\subsection{Commit Message Generation}

Researchers have proposed several approaches for commit message generation. According to different generating mechanisms, they can be categorized into retrieval-based, learning-based, and hybrid methods. 

\subsubsection{Retrieval-based methods} NNGen~\cite{nngen} leverages information retrieval techniques to suggest commit messages from similar code diffs. To generate a commit message, NNGen calculates the cosine similarity between the target code diff and each code diff in the collected corpus. Then, the top-k diff-message pairs are selected to compute the BLEU scores, the one with the highest score is regarded as the most similar code diff, and its commit message will be used as the target one. CC2Vec~\cite{cc2vec} learns a representation of code diff guided by their accompanying commit messages. Similar to the nearest neighbors approach, it computes the distance between code diff vectors and directly outputs the commit message of the closest CC2Vec vector. Retrieval-based methods have significant challenges as the corpus is limited and cannot cover all code diffs.

\subsubsection{Learning-based methods} Learning-based methods~\cite{cmtgen,nmt,codisum,fira,ptrgncmsg} typically employ deep learning techniques, treating CMG as a Neural Machine Translation (NMT) task. These methods learn how to generate commit messages by training deep neural network models on massive diff-message datasets collected from GitHub projects. CommitGen~\cite{cmtgen} is an early attempt to use NMT in CMG, it trained a recurrent neural network (RNN) encoder-decoder model using a corpus of diffs and human-written commit messages from the top 1k GitHub projects. CoDiSum~\cite{codisum} uses a multi-layer bidirectional gated recurrent unit (GRU) as its encoder part, which can better learn the representations of code changes. Moreover, the copying mechanism is used in the decoder part to mitigate the out-of-vocabulary (OOV) issue. PtrGNCMsg~\cite{ptrgncmsg} is another NMT approach based on an improved sequence-to-sequence model with the pointer-generator network which is an adapted version of the attention RNN encoder-decoder model. The most recently proposed learning-based approach is FIRA~\cite{fira}. It first represents the code diffs with fine-grained graphs, which explicitly describe the code edit operations between the old version and the new version, and code tokens at different granularities. The hybrid architecture of transformer and GNN is adopted as the backbone of the model. FIRA outperforms other learning-based methods and can be considered the current state-of-the-art (SOTA) approach of a single model.

\subsubsection{Hybrid methods} ATOM~\cite{ATOM} is a hybrid method containing three modules, a generation module encoding the structure of code diff using Abstract Syntax Tree (AST), a retrieval module retrieving the most similar message based on the text-similarity, and a hybrid ranking module selecting the best commit message from the ones generated by generation and retrieval modules. CoRec~\cite{corec} takes advantage of both IR and NMT, addressing the low-frequency word and exposure bias issue. It trained a context-aware encoder-decoder model. Given a diff for testing, the method retrieves the most similar diff from the training set and then uses it to guide the probability distribution for the final generated vocabulary. RACE~\cite{race} combines retrieval and generation techniques in a more integrated way but employs a trivial retriever and opts for training a specific model from scratch. Whereas, our proposed framework REACT incorporates a hybrid advanced retriever, and experimental results demonstrate its effectiveness. Furthermore, REACT is not limited to a specific model, but rather widely adopts various successful models as the generator, enabling more effective generation of commit messages and providing generality insights.

\subsection{Code Language Models}
Code Language Models (CLMs) refer to language models trained on datasets containing code, equipping them with the capability to generate and understand code. This makes CLMs applicable for downstream code tasks, including but not limited to code completion, program repair, code summarization, and others. In this work, we present a comprehensive evaluation of CLMs on the CMG task, and CLMs act as the generator in our proposed framework. Based on their scale and the methods for transferring to downstream tasks, they can be categorized into the following two types:

\subsubsection{Pre-trained Language Models (PLMs) for Code} PLMs are the language models that have been pre-trained on massive training datasets but require fine-tuning when transferred to downstream tasks. Researchers have recently introduced numerous impressive PLMs for code~\cite{codebert,codet5,codet5p,unixcoder,plbart}. For instance, CodeT5~\cite{codet5,codet5p} is a unified pre-trained encoder-decoder transformer~\cite{vaswani2017attention} model with the identifier-aware pre-training objective on large-scale program language and natural language datasets, which has impressive code understanding and generation capabilities. There are some empirical studies of PLMs for various tasks in the software engineering field~\cite{plm4apr, liu2020multi, xia2022less, lu2021codexglue}, showcasing their remarkable success in code-related tasks.

\subsubsection{Large Language Models (LLMs)} The emergence of ChatGPT marks the beginning of the LLMs era. LLMs possess massive numbers of parameters (exceeding billions), which are typically instruction-tuned~\cite{instruct}, enabling them to be directly applied to downstream tasks without training by simply designing the instruct prompt. LLMs generally have code capabilities, and code-related tasks are important parts of LLM benchmarks. Meta recently released its latest series of open-source LLMs, Llama 3\footnote{\url{https://llama.meta.com/llama3/}}. Google has also released its open-source LLM, Gemma\footnote{\url{https://ai.google.dev/gemma}}, which is featured with lightweight. These LLMs have contributed a lot to the open-source community, benefiting countless research activities.
\begin{figure*}[t]
    \centering
    \includegraphics[width=\linewidth]{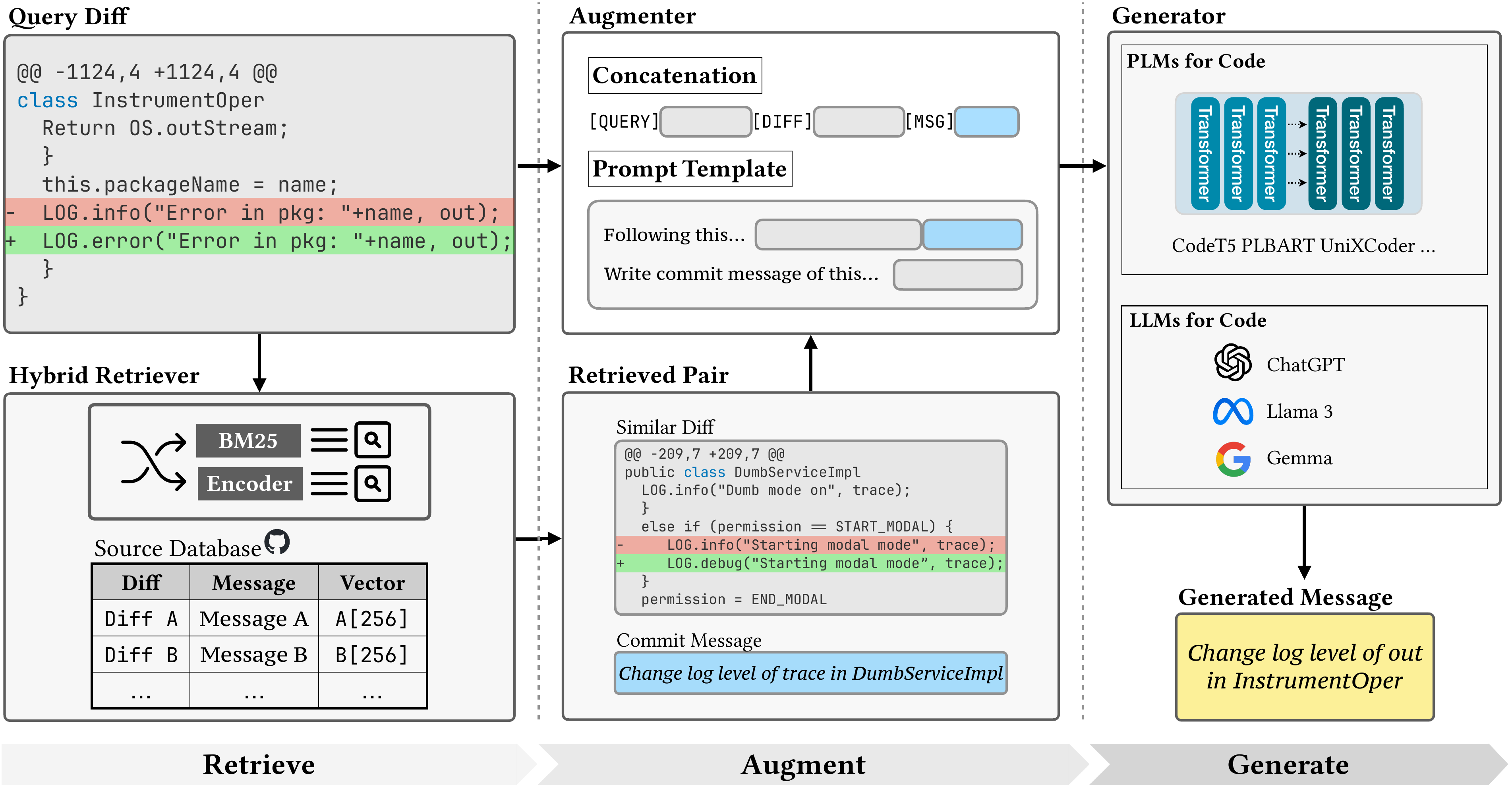}
    \caption{Overview of our REACT framework. Regarding a code diff (query diff) awaiting the generation of its corresponding commit message, we first retrieve a relevant diff-message pair from the source database using the hybrid retriever. After that, we put query diff and retrieved pair together into the augmenter for concatenation or filling the prompt template. The alterable generator receives formatted inputs and generates a commit message under the guidance of the retrieved pair. }
    \label{fig:overview}
\end{figure*}
\subsection{Retrieval-Augmented Generation}

Retrieval-augmented generation~\cite{rag} involves enhancing the generation of existing models by incorporating knowledge from external databases. A common paradigm includes three steps: information retrieval, data augmentation, and final generation. It has been widely applied to various tasks in NLP, particularly benefiting knowledge-intensive tasks by allowing for the integration of domain-specific information, and has achieved impressive performance. Recent works have also applied this paradigm of RAG to code-related tasks, including code summarization~\cite{codesumm, codesumm2}, program repair~\cite{rapgen}, and code completion~\cite{lu-etal-2022-reacc}.

\section{Methodology}
\label{sec:3}

This section presents the methodology of our proposed three-phrase framework, i.e., REACT. We first provide a brief overview of REACT. Then, the three phases of REACT are addressed in detail. 

\subsection{Overview}
As shown in Fig.~\ref{fig:overview}, REACT comprises three phases, i.e., \textit{Retrieve}, \textit{Augment}, and \textit{Generate}, to generate the commit message for a specified code diff. In this paper, we refer to the code diff that awaits the generation of its corresponding commit message as the ``query diff''. Firstly, REACT retrieves the most relevant diff-message pair from the source database using a hybrid retriever. Then, the query diff and the retrieved diff-message pair are sent together to the augmenter for augmentation. Finally, the augmented input is fed into the generator to generate a commit message under the guidance of the retrieved pair.


\subsection{Phase \uppercase\expandafter{\romannumeral1}: Retrieve}
\label{sec:retrieve}
Phase \uppercase\expandafter{\romannumeral1} intends to retrieve the most relevant diff-message pair for guidance. which  is accomplished through the following two components:

\subsubsection{Hybrid Retriever}

To more accurately assess the similarity between two code diffs and retrieve the most relevant diff-message pair based on that similarity, we designed an advanced hybrid retriever that involves combining two similarity scores with weighted fusion. 
\textbf{BM25}~\cite{bm25} is a relevance scoring algorithm commonly used in information retrieval and search engines. It measures the relevance of a document to a given query by considering factors such as term frequency, document length, and term saturation. BM25 treats query diff as a bag-of-words representation and computes lexical similarity scores between query diff and each of the candidates. 
We also employ a transformer \textbf{encoder}-based similarity scoring method. As an improved version of CodeT5~\cite{codet5}, CodeT5+~\cite{codet5p} is a cutting-edge PLM for code with the architecture of transformer encoder-decoder. Through its pre-training process, it learned rich representations from code data. We extract the encoder component of CodeT5+ and then utilize it to transform the code diff text into a dense 256-dimensional vector embedding, which is generally considered to encapsulate the semantic information of the code diff. By calculating the cosine similarity score between two vectors, we can obtain the similarity score between two code diffs. Following the normalization of the scores to a common scale, we combine the scores obtained from these two methods with equal weights (1:1) in our experiment, and we refer to it as the \textit{hybrid score}. After calculating the hybrid scores between the query diff and all diffs in the source database, the diff-message pair with the highest score is retrieved as the most relevant pair. The hybrid retriever is expected to be more robust and effective compared to single retrievers that rely on only one similarity scoring method.

\textit{Data leakage issue:} It should be pointed out that due to the intersection between the test set and the source database in the actual experiment, the retrieved diff-message pair may be exactly the same as that of the test set, resulting in data leakage issues. We append a simple mechanism to avoid this issue: when the retrieved diff is detected to be the same as the query diff, the pair with the second-highest hybrid score is selected to replace it. 

\subsubsection{Source Database}
To ensure that each query diff can retrieve a relevant diff-message pair as an exemplar and that the retrieved exemplar is of high quality, containing sufficient information to guide generation, a comprehensive, high-quality source database is crucial. To avoid reinventing the wheel, we reuse the data from CommitBench~\cite{commitbench}, a recently proposed dataset, to construct the source database used in our work. CommitBench collects commits from 72,000 publicly available GitHub repositories, which is the largest collection scope among existing datasets, ensuring the comprehensiveness of the dataset. Regarding the quality of this dataset, the repositories collected by CommitBench are non-fork repositories, which have been used by at least one other project and have a reasonable number of stars. Additionally, they performed an overall filtering process, designing ten filtering rules to exclude worthless messages, including bot messages, too-short or too-long messages, etc. The resulting source database contains over 1.6 million diff-message pairs. During the pre-processing stage, we persisted the 256-dimensional vector embedding of code diffs, obtained through the encoder in the hybrid retriever, as a new column in the source database to avoid additional computational overhead.

\subsection{Phase \uppercase\expandafter{\romannumeral2}: Augment}
\label{sec:augment}

This phase aims at combining the query diff and the retrieved pair into specific formats to augment the input context. For ease of reference, we denote \texttt{<query diff>}, \texttt{<retrieved diff>}, \texttt{<retrieved msg>} as the query diff for generation, the retrieved code diff, and the corresponding retrieved commit message, respectively. Considering the differences between PLMs and LLMs, we design the following two input augmentation methods for both:

\subsubsection{For PLMs} PLMs require a specific input format to serve as the generator in REACT. To explicitly distinguish between different input components, we define three special tokens, i.e., \texttt{[QUERY]}, \texttt{[DIFF]}, and \texttt{[MSG]}. The augmenter concatenates these special tokens with query diff and retrieved pair into the following form:

{
\small \begin{verbatim}
[QUERY]<query diff>[DIFF]<retrieved diff>[MSG]<retrieved msg>
\end{verbatim}
}

These three special tokens will be added to tokenizers of PLMs and will be fully trained during the fine-tuning stage.

\subsubsection{For LLMs} 
\begin{figure}[b]
    \centering
    \includegraphics[width=0.6\linewidth]{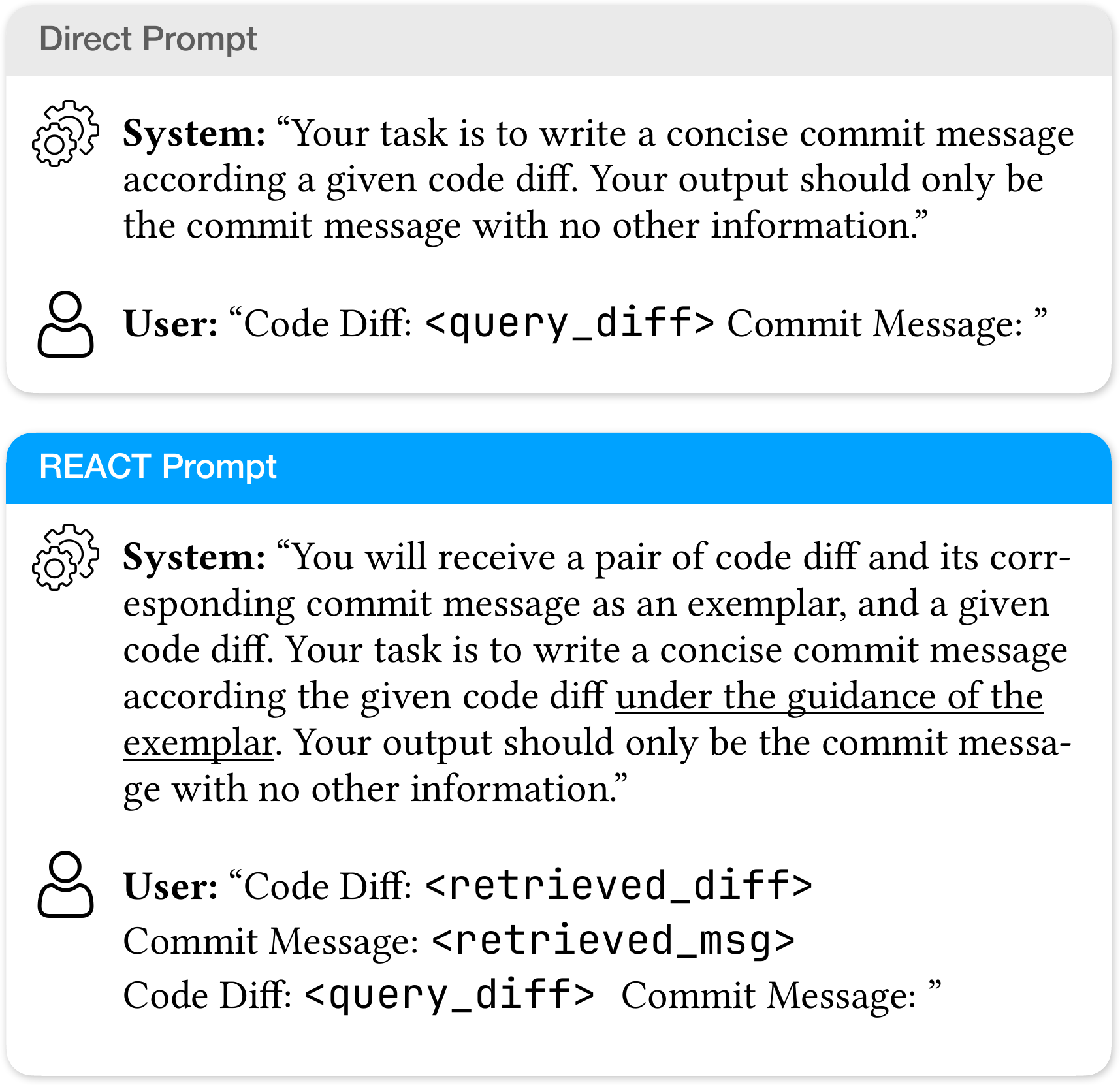}
    \caption{Direct prompt template and REACT prompt template for LLMs.}
    \label{fig:prompt}
\end{figure}

We manually construct a prompt template for LLMs, as Figure~\ref{fig:prompt} shows, to guide commit message generation by referring to retrieved exemplar. The augmenter fills in the corresponding content into the prompt template and then inputs it into LLMs for generation.

\subsection{Phase \uppercase\expandafter{\romannumeral3}: Generate} 

The generator receives the augmented input from Phase \uppercase\expandafter{\romannumeral2} and generates the final commit message under the guidance of the retrieved pair inserted in the input. In this paper, we adopt various models, including PLMs and LLMs, as the generators and conduct a comprehensive study. The specific model selection will be detailed in Section~\ref{sec:model}. In this subsection, we discuss the two types of generators used in REACT and what adjustments they need to make to perform generating.

\subsubsection{PLMs} \label{sec:ft} PLMs for code have impressive code understanding capabilities and are well suited for the CMG task. As pre-trained models, PLMs require fine-tuning to fit our generating needs. Fine-tuning is essentially a model training process where PLMs are expected to learn how to generate better commit messages based on the additional information provided by the exemplar. 

We denote the augmented input to the PLMs generator as $X$: {\small\texttt{[QUERY]<query diff>[DIFF]<retrieved diff>[MSG]<retrieved msg>}}, and the expected output (ground truth) as $\hat{Y}$, that is, commit messages written by the developer. The objective of fine-tuning is to minimize the cross entropy~\cite{shore1981cross} loss $\mathcal{L}(X, \hat{Y})$ of all instances of the training set defined as:

$$
\mathcal{L}(X, \hat{Y}) = -\sum_{t=1}^{T} \log P_\theta(\hat{Y}_t | X)
$$

where $P_\theta(\hat{Y}_t | X)$ represents the probability assigned by the PLMs to the correct token $\hat{Y}_t$ at time step $t$.
After adequate training, the overall parameters of PLMs are updated, and they can be used as the generator in REACT. 
\subsubsection{LLMs} In contrast to PLMs, LLMs can serve as the generator to generate commit messages without training. It benefits from the powerful generalization ability of LLMs, which can complete the specified tasks with only prompting~\cite{instruct}. Meanwhile, through the well-designed instructions as shown in Figure~\ref{fig:prompt}, LLMs can perform generating under the guidance of exemplars with in-context learning~\cite{incontext}. By providing a relevant exemplar in the input, that is, one-shot prompting~\cite{brown2020language}, LLMs can perform the CMG task with reference to the provided exemplar. The more relevant the provided exemplar is to the query diff, the more valuable it is as a reference, and LLMs are expected to generate better commit messages.

\section{Experimental Setup}
\label{sec:4}

In this section, we first propose four research questions that reflect the basis of our experimental path. Then, we introduce the dataset used in the experiments, the selected models, the baselines for comparison, the metrics used for evaluating results, as well as the implementation details for reference and reproduction.

\subsection{Research Questions}
The overall goal of this research is to validate the effectiveness of REACT.
To achieve this goal, we conducted experiments aimed at investigating the following five research questions (RQs):

\textbf{RQ1: What is the efficacy of \textit{directly} applying CLMs to the CMG task?} How do CLMs perform on the CMG task when directly fine-tuned or prompted with basic instructions, and how do they compare to existing baselines? This RQ is designed to explore the applicability of chosen CLMs on the CMG task. 

\textbf{RQ2: Does REACT enhance the performance of PLMs on the CMG task?} When PLMs are integrated into the REACT framework as the generator, to what extent does REACT enhance the performance of PLMs on the CMG task? This RQ is designed to investigate the effectiveness of REACT when integrating with PLMs. 

\textbf{RQ3: Does REACT enhance the performance of LLMs on the CMG task?} To what extent does REACT enhance the performance of LLMs on CMG when LLMs are integrated into the framework as the generator and augmented with a distinct in-context learning paradigm? This RQ is designed to investigate the effectiveness of REACT when integrated with LLMs, providing generality insights.

\textbf{RQ4: Can our hybrid retriever find relevant diff-message pairs to guide generation?} Compared to using single retrieval techniques, can the hybrid retriever better retrieve relevant pairs to guide generation? This RQ aims to investigate the ablation study of the hybrid retriever.

\textbf{RQ5: Can REACT perform well in the within-project scenario?} Considering a within-project scenario where commit messages typically follow a consistent style, can REACT adapt its generated commit messages to align with the conventions of the specific project, enhancing both the relevance and usability of the output? This RQ is designed to investigate REACT’s ability to generate project-specific commit messages by leveraging historical records from the same project.

\subsection{Dataset}
In this work, we employ a well-established dataset~\cite{codisum} that has been widely used in previous CMG works~\cite{cc2vec, codisum, fira, cmtgen, nngen,corec}. The dataset contains a total of 90,661 pairs of code diff and corresponding commit messages collected from the top 1,000 popular Java repositories in GitHub. Following the previous work~\cite{codisum}, the dataset is divided into training, validation, and testing sets in the ratio, as shown in Table~\ref{tab:datasets}. In addition to using the original dataset to evaluate the performance of directly applying CLMs to the CMG task in RQ1, we also construct an augmented dataset. The augmented dataset is derived from the original dataset through a retrieval augmentation process described in Section~\ref{sec:retrieve} and Section~\ref{sec:augment}. Thus, for each query diff in the original dataset, a relevant diff-message pair is retrieved and saved. The augmented dataset is used for assessing the effectiveness of REACT in RQ2--RQ4. Table~\ref{tab:datasets} presents the statistics of both the original and augmented datasets, in which ``Count'' refers to the number of entries in each dataset split. We find that since each entry of the augmented dataset includes two diff-message pairs (query and retrieved pair), the number of tokens nearly doubles. 
All the models in the experiments are evaluated on the testing set. The training and validation sets are used in the fine-tuning stage of PLMs.

\begin{table}[t]
  \caption{Statistics of Original and Augmented Datasets.}
  \label{tab:datasets}
  \begin{tabular}{l|cc}
    \toprule
    \textbf{Dataset} & \textbf{Origin. (RQ1)} & \textbf{Augmen. (RQ2--4)} \\
    \midrule
    Count in Training Set    & 75,000 & 75,000 \\
    Count in Validation Set  & 8,000  & 8,000  \\
    Count in Testing Set     & 7,661  & 7,661  \\
    Avg. tokens of Message   & 9.1    & 28.4   \\
    Avg. tokens of Diff      & 338.8  & 634.5  \\
    \bottomrule
\end{tabular}
\end{table}

\begin{table}[t]
  \caption{Selected Models.}
  \label{tab:model}
  \begin{tabular}{cl|c|c}
    \toprule
    ~ & \textbf{Model}&\textbf{Parameters}&\textbf{Architecture}\\
    \midrule
    \multirow{4}{*}{\textit{PLMs}} & CodeT5~\cite{codet5} & 220M & \multirow{4}{*}{Encoder-Decoder}\\
    ~ & CodeT5+~\cite{codet5p} & 220M \\
    ~ & PLBART~\cite{plbart} & 140M \\
    ~ & UniXCoder~\cite{unixcoder} & 126M \\
    \midrule
    \multirow{3}{*}{\textit{LLMs}} & ChatGPT~\cite{chatgpt} & N/A & \multirow{3}{*}{Decoder-only}\\
    ~ & Llama 3~\cite{llama3} & 70B & ~\\
    ~ & Gemma~\cite{gemma} & 7B & ~\\
  \bottomrule
\end{tabular}
\end{table}

\subsection{Model Selection}
\label{sec:model}

Table~\ref{tab:model} shows the selected CLMs in this paper. Specifically, in the experiment, we adopt the bimodal version for PLMs to acquire the capacities of both natural language and code. For ChatGPT, we use the GPT-4o API version. For Llama 3 and Gemma, we adopt the instruct version so that LLMs can follow the prompt instructions.

In our experiments, in addition to serving these models as the generator in REACT, we also evaluate the performance of these models directly applied to the CMG task for comparison, in which ``directly applying'' means using the original dataset without augmentation to generate commit messages with no guidance of exemplars.

\subsection{Compared Baselines}
We choose the following six existing CMG methods as baselines:

Retrieval-based methods: We consider the representative retrieval based method \textbf{NNGen}~\cite{nngen}, which is the first work to apply IR techniques to the CMG task.

Learning-based methods: We consider three learning-based methods leveraging deep learning models for CMG, i.e., \textbf{CommitGen} \cite{cmtgen}, \textbf{PtrGNCMsg}~\cite{ptrgncmsg}, and \textbf{FIRA}~\cite{fira}.

Hybrid methods: We consider two hybrid methods that take advantage of both IR and NMT techniques. They are \textbf{CoRec}~\cite{corec} and \textbf{RACE} \cite{race}.

\subsection{Metrics}
Following previous works on CMG ~\cite{cc2vec, codisum, fira}, we employ three widely used metrics, \textbf{BLEU}, \textbf{Rouge-L}, and \textbf{METEOR} to measure the similarity. The evaluation metrics compute similarity scores between the generated text and the ground truth, higher is better.
\begin{itemize}
    \item \textbf{BLEU}~\cite{bleu} is a metric originally used in machine translation evaluation. In this task, it is computed based on the similarity of n-gram between the generated commit message and ground truth (developer-written message). The n-gram precision refers to the ratio of the number of matched n-grams to the number of all the n-grams in the generated text. We use BLEU-4 with 4-grams in the experiments.
    \item \textbf{Rouge-L}~\cite{rouge} stands for Recall-Oriented Understudy for Gisting Evaluation. The measures count the number of overlapping units such as n-grams, word sequences, and word pairs. Rouge-L regards the longest common subsequence (LCS) between the two sentences.
    \item \textbf{METEOR}~\cite{meteor} is calculated based on the harmonic mean of precision and recall, with recall weighted more than precision. It is based on a generalized concept of unigram matching between the generated text and ground truth. It is also widely used in machine translation evaluation.
\end{itemize}

For simplicity and consistency, all the metrics use the implementation provided by HuggingFace\footnote{\url{https://huggingface.co/evaluate-metric}}.

\subsection{Implementation Details}

The PLMs used in this study are implemented from the official HuggingFace repository\footnote{\url{https://huggingface.co/Salesforce/codet5-base}}\footnote{\url{https://huggingface.co/Salesforce/codet5p-220m-bimodal}}\footnote{\url{https://huggingface.co/uclanlp/plbart-base}}\footnote{\url{https://huggingface.co/microsoft/unixcoder-base}}, Gemma and Llama 3 are based on the official released model weights and use FP16 precision. ChatGPT used the GPT-4o API as of the experiment date, Aug 1. The fine-tuning of the PLMs was completed on an NVIDIA RTX 4090, utilizing the Adam optimizer to minimize the cross-entropy loss function. The models were trained for 20 epochs using a batch size of 32 and a learning rate of 5e-5, and the same hyperparameters were applied to all four PLMs.
Due to the vast size of the source database, which contains millions of entries, we employed the industrial-grade search engine Apache PyLucene~\cite{lucene} to calculate BM25 scores. This ensures that the retrieval average time for one query remains acceptably short.

\section{Results and Analysis}
\label{sec:5}

In this section, we present and analysis the experimental results to answer the proposed four RQs.

\begin{table}[b]
  \caption{Results of directly using CLMs on the CMG task.}
  \label{tab:rq1}
  \begin{tabular}{c l c c c}
    \toprule
    ~ & \multirow{2}{*}{\textbf{Model}} & \multicolumn{3}{c}{\textbf{Metric Scores (\%)}}  \\
    \cmidrule{3-5}
    ~ & ~ & BLEU & Rouge-L & METEOR \\
    \midrule
    \midrule
    \multirow{6}{*}{\begin{sideways}\small\textit{Baselines}\end{sideways}} & CommitGen~\cite{cmtgen} & 1.36 & 10.59 & 9.17 \\
    ~ & NNGen~\cite{nngen} & 3.09 & 10.54 & 10.11 \\
    ~ & CoRec~\cite{corec} & 3.03 & 12.28 & 10.75 \\
    ~ & PtrGNCMsg~\cite{ptrgncmsg} & 0.99 & 16.06 & 11.89 \\
    ~ & FIRA~\cite{fira} & 2.80 & 21.56 & 14.75 \\
    ~ & RACE~\cite{race} & \cellcolor{gray!7}5.16 & \cellcolor{gray!20}24.72 & 18.19 \\
    \midrule
    \multirow{3}{*}{\begin{sideways}\small\textit{LLMs}\end{sideways}} & Gemma~\cite{gemma} & 1.54 & 13.50 & 15.04 \\
    ~ & Llama 3~\cite{llama3} & 2.40 & 17.91 & \cellcolor{gray!7}18.65\\
    ~ & GPT-4o~\cite{chatgpt} & 2.57 & 20.25 & \cellcolor{gray!13}19.03 \\
    \midrule
    \multirow{4}{*}{\begin{sideways}\small\textit{PLMs}\end{sideways}} & UniXCoder~\cite{unixcoder} & \cellcolor{gray!20}5.24 & \cellcolor{gray!13}23.88 & 17.95 \\
    ~ & PLBART~\cite{plbart} & \cellcolor{gray!13}5.17 & \cellcolor{gray!7}23.30 & \cellcolor{gray!20}19.90 \\
    ~ & CodeT5+~\cite{codet5p} & \cellcolor{gray!30}6.00 & \cellcolor{gray!30}25.66 & \cellcolor{gray!40}\textbf{21.75} \\
    ~ & CodeT5~\cite{codet5} & \cellcolor{gray!40}\textbf{6.24} & \cellcolor{gray!40}\textbf{25.85} & \cellcolor{gray!30}21.71 \\
  \bottomrule
\end{tabular}
\end{table}

\subsection{RQ1: Efficacy of Directly Applying CLMs to the CMG Task} 
\label{sec:rq1}
At the beginning, we evaluated the CLMs selected for the experiments by directly applying them to the CMG task to assess their feasibility. These CLMs have already been widely applied to various code-related tasks, such as code completion, code summarization, and automated program repair. Additionally, some recent works~\cite{zhang2024using,llmbeyond,lopes2024commitllm} have explored the performance of some CLMs, e.g. CodeBERT, Llama, and ChatGPT, on the CMG task.
However, the evaluations of these studies were conducted on small sampled datasets, and their model selections were neither comprehensive coverage nor cutting-edge.
Therefore, it is necessary to comprehensively evaluate and compare the performance of selected CLMs directly applied to the CMG task before investigating the remaining RQs.

 To answer RQ1, we selected seven models, covering both PLMs and LLMs and including the latest models (e.g., Llama 3 released in April 2024). Moreover, the evaluations of these seven models are conducted on the whole dataset and the results are compared with existing SOTA baselines. 

\begin{figure}[t]
    \centering
    \includegraphics[width=0.6\linewidth]{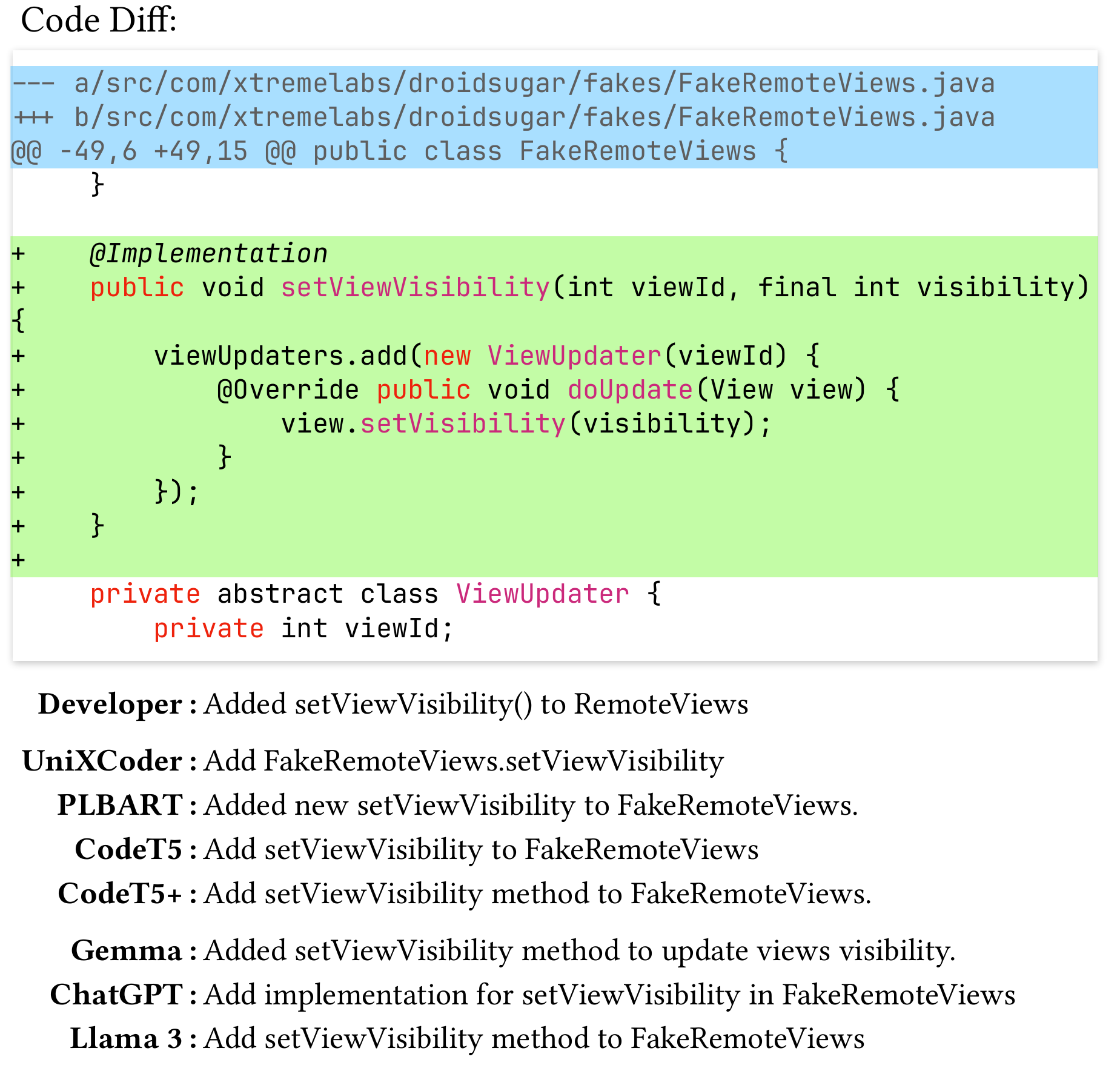}
    \caption{An example of CLMs generating commit message according to a code diff.}
    \label{fig:case}
\end{figure}

\subsubsection{Results}

Table~\ref{tab:rq1} lists the scores of CLMs and baselines on the testing set across three metrics (higher is better). The overall results show that CodeT5 performs the best, achieving the highest scores in both BLEU and Rouge-L metrics and significantly surpassing the baselines. Among the four PLMs, three models ranked in the top-3, demonstrating a clear advantage. LLMs obtain decent scores, with METEOR scores matching or exceeding the baselines. However, for the other two metrics, they fall behind FIRA and RACE. Figure~\ref{fig:case} shows an example of code diff and its corresponding commit messages generated by selected CLMs.

\subsubsection{Analysis}

For PLMs, it is impressive that they achieved a significant lead with only simple fine-tuning. Specifically, all three metric scores of CodeT5 are higher than that of RACE, which performs the best in the baselines. Moreover, to achieve such performance, CodeT5 required only 8 hours of training on a single RTX 4090 GPU, whereas RACE took 14 hours in the same environment. The fine-tuning paradigm of PLMs for code not only achieves better performance but also requires relatively fewer resources. In fact, this is the superiority of PLMs. They have undergone extensive pre-training on massive datasets, acquiring sufficient prior knowledge. This allows them can be easily transferred to downstream tasks including CMG through fine-tuning. On the other hand, existing CMG methods generally choose to train a model from scratch, which limits their performance. Compared to teaching a child who is just learning to speak (an initial weight model) how to write commit messages, it is easier and more effective to teach a programmer who already knows code and language (a pre-trained model like CodeT5) to do so. 

For LLMs, although they do not perform as well as RACE in terms of BLEU and Rouge-L scores, they are equivalent to other baselines. In particular, Llama 3 surpasses all the baselines in the METEOR score. It is worth noting that the LLMs were directly applied to the CMG task without any training or fine-tuning, achieving such results through simple prompting in an out-of-the-box manner. This performance is quite remarkable and also proves that LLMs have sufficient usability for the CMG task.

Overall, the results of RQ1 demonstrate the strong capabilities of CLMs and their suitability for the CMG task. This is owing to the rich prior knowledge embedded in CLMs, which establishes a foundation for their subsequent use as the generator in the REACT framework.

\begin{framed}
    \noindent \textbf{Key Findings of RQ1:} 
    \begin{itemize}
        \item When directly applying to the CMG task, PLMs significantly outperformed the baselines across the board.
        \item While LLMs lagged behind the best baseline model in BLEU and Rouge-L metrics, Llama 3 surpassed all baselines in the METEOR score.
        \item These results demonstrate the superiority of CLMs, attributed to their prior knowledge, which enables them to be effectively transferred to the CMG task through simple fine-tuning or prompting.
    \end{itemize}
\end{framed}

\begin{table}[b]
  \caption{Results of REACT integrating with CLMs. ``$\uparrow$X / $\downarrow$X'': relative changes over directly applying.}
  \label{tab:rq2}
  \begin{tabular}{cl|ccc}
  \toprule
    \multicolumn{2}{c}{\multirow{2}{*}{\textbf{Approach}}} & \multicolumn{3}{c}{\textbf{Metric Scores (\%)}}  \\
    \cmidrule{3-5}
    ~ & ~ & BLEU & Rouge-L & METEOR \\
    \midrule
    \midrule
    \multirow{2}{*}{UniXCoder} & \textit{directly} & 5.24 & 23.88 & 17.95\\
    ~ & \textit{REACT} & \textbf{9.25} {$\uparrow$\textbf{76\%}} & \textbf{25.66} {$\uparrow$\textbf{7.5\%}} & \textbf{20.16} {$\uparrow$\textbf{12\%}}\\
    \cmidrule{1-5}
    \multirow{2}{*}{PLBART} & \textit{directly} & 5.17 & 23.30 & 19.90\\
    ~ & \textit{REACT} & \textbf{6.95} {$\uparrow$\textbf{34\%}} & \textbf{24.15} {$\uparrow$\textbf{3.6\%}} & 19.25 {$\downarrow$3.3\%} \\
    \cmidrule{1-5}
    \multirow{2}{*}{CodeT5} & \textit{directly} & 6.24 & 25.85 & 21.71 \\
    ~ & \textit{REACT} & \textbf{9.68} {$\uparrow$\textbf{55\%}} & \textbf{27.20} {$\uparrow$\textbf{5.2\%}} & \textbf{23.65} {$\uparrow$\textbf{8.9\%}} \\
    \cmidrule{1-5}
    \multirow{2}{*}{CodeT5+} & \textit{directly} & 6.00 & 25.66 & 21.75 \\
    ~ & \textit{REACT} & \textbf{9.59} {$\uparrow$\textbf{60\%}} & \textbf{26.30} {$\uparrow$\textbf{2.5\%}} & \textbf{22.88} {$\uparrow$\textbf{5.2\%}} \\
    \cmidrule{1-5}
    \multirow{2}{*}{Gemma} & \textit{directly} & 1.54 & 13.50 & 15.04 \\
    ~ & \textit{REACT} & \textbf{2.32} {$\uparrow$\textbf{51\%}} & 9.69 {$\downarrow$28\%} & 11.64 {$\downarrow$23\%} \\
    \cmidrule{1-5}
    \multirow{2}{*}{GPT-4o} & \textit{directly} & 2.57 & 20.25 & 19.03 \\
    ~ & \textit{REACT} & \textbf{3.89} {$\uparrow$\textbf{51\%}} & 20.45 $\sim$0\% & \textbf{19.78} {$\uparrow$\textbf{3.9\%}} \\
    \cmidrule{1-5}
    \multirow{2}{*}{Llama 3} & \textit{directly} & 2.40 & 17.91 & 18.65 \\
    ~ & \textit{REACT} & \textbf{4.84} {$\uparrow$\textbf{102\%}} & \textbf{19.88} {$\uparrow$\textbf{11\%}} & \textbf{20.58} {$\uparrow$\textbf{10\%}} \\
  \bottomrule
\end{tabular}
\end{table}

\subsection{RQ2: Effectiveness of REACT When Integrating with PLMs}

For clarity, RQ2 and RQ3 are separated into distinct two research questions because integrating PLMs or LLMs into the REACT framework corresponds to the two different paradigms: fine-tuning or in-context learning. Therefore, we investigate them separately in the two RQs. 
This RQ primarily explores the effectiveness and extent to which the REACT framework can enhance the CMG performance of PLMs. Before integrating PLMs into the REACT framework as the generator, they must undergo fine-tuning as described in section~\ref{sec:ft}. The experiment involved evaluating four PLMs across the entire testing set.

\subsubsection{Results} The first four rows of Table~\ref{tab:rq2} show the comparison results between directly applying PLMs and integrating PLMs into the REACT framework. The results indicate that REACT enhances the performance of all four PLMs evaluated by a large margin. Specifically, when CodeT5 is integrated as the generator within REACT, it achieves the highest BLEU score of 9.68, representing a 55\% improvement compared to using CodeT5 directly. Its Rouge-L and METEOR scores also increase by 5.2\% and 8.9\%, respectively. More importantly, integrating CodeT5 into the REACT framework results in metric scores that surpass all baselines, establishing a new state-of-the-art (SOTA). For the other three PLMs, REACT also provides varying degrees of improvement. For UniXCoder, the BLEU score shows a maximum percentage increase of up to 76\%. Overall, integrating PLMs into REACT can significantly increase BLEU scores, the percentage increases in Rouge-L and METEOR scores are more modest but still notable.

\subsubsection{Analysis} The substantial enhancement in metric scores suggests that REACT effectively leverages the prior knowledge embedded in PLMs, allowing them to generate more accurate and relevant commit messages under the guidance of retrieved exemplars. The results also reveal that the integration of retrieval-augmented techniques is particularly beneficial for PLMs with strong pre-existing capabilities, like CodeT5 and UniXCoder. These models showed the most substantial gains. Whereas the improvement of PLBART is relatively small.

\begin{framed}
    \noindent \textbf{Key Findings of RQ2:} 
    \begin{itemize}
        \item Integrating PLMs into the REACT framework significantly enhance their CMG performance.
        \item With REACT, UniXCoder achieves the highest BLEU percentage increase of 76\%, while CodeT5 attains the highest BLEU score of 9.68, establishing a new SOTA.
    \end{itemize}
\end{framed}

\subsection{RQ3: Effectiveness of REACT When Integrating with LLMs}

Integrating LLMs into the REACT framework requires no additional training. Compared to direct application, the prompt input to LLMs in REACT includes an example diff-message pair as shown in Figure~\ref{fig:prompt}. This essentially transforms a zero-shot scenario into a one-shot scenario by incorporating an exemplar into the context of the LLMs' input, with the expectation that this will enable the LLMs to generate better commit messages.
\begin{figure}[b]
    \centering
    \includegraphics[width=\linewidth]{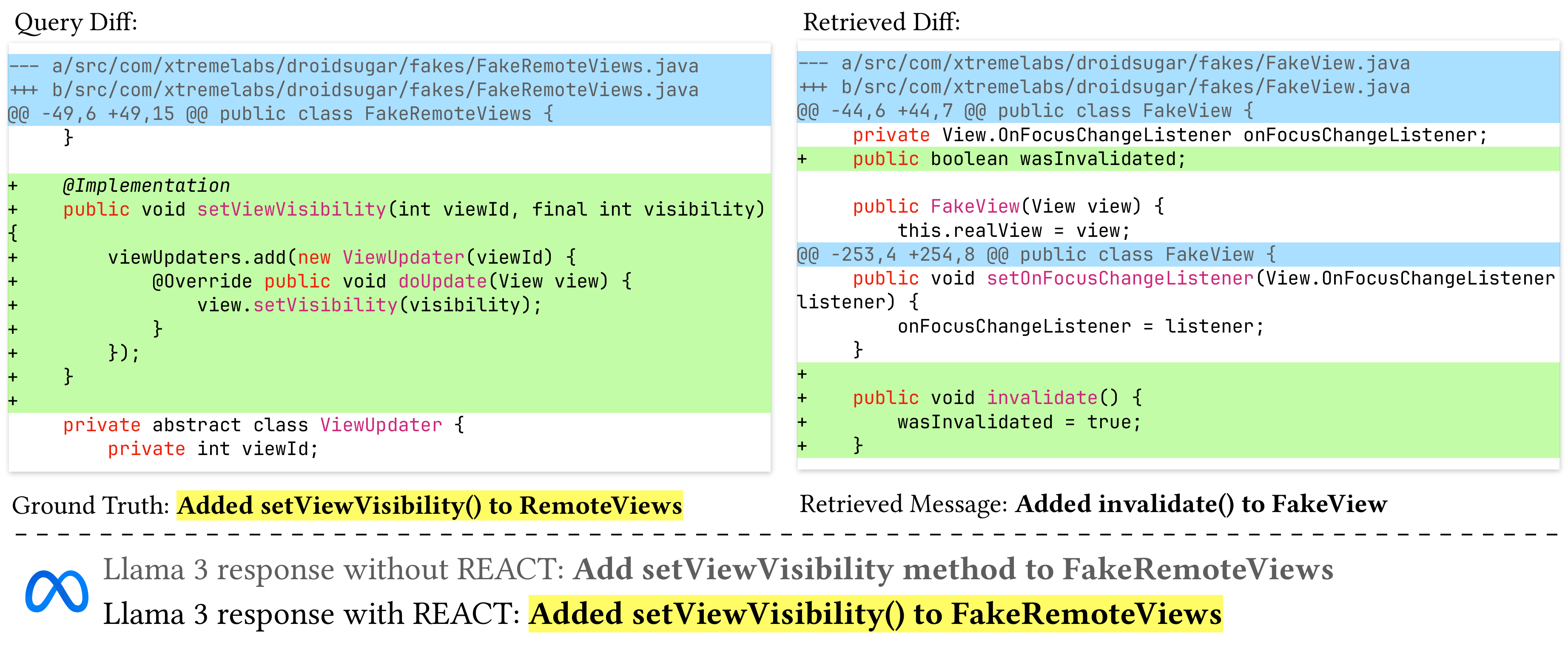}
    \caption{A case of generating commit message using Llama 3 integrated with REACT.}
    \label{fig:case2}
\end{figure}
\subsubsection{Results}

The last three rows of Table~\ref{tab:rq2} show the comparison results between directly applying LLMs and integrating LLMs into the REACT framework. From the perspective of BLEU scores, REACT significantly and consistently improves LLM performance. Specifically, Llama 3's BLEU score increased from 2.40 to 4.84, representing a 102\% improvement. However, for the Rouge-L and METEOR scores, while both ChatGPT and Llama 3 show varying degrees of improvement, Gemma exhibits a noticeable decrease and its scores are comparatively lower than the other two LLMs. Figure~\ref{fig:case2} shows a case using REACT, the query diff is the same as the one of Figure~\ref{fig:case}, with the relevant exemplar shown on the right. We can see that both commits originate from different files within the same repository, exhibiting similar commit message writing styles. By referencing the exemplar, Llama 3 with REACT can effectively adopt this writing style,  resulting in commit messages that closely align with those written by developers, thereby improving the metric scores.

\subsubsection{Analysis}

REACT can enhance the performance of ChatGPT and Llama 3 in CMG. However, the results for Gemma are somewhat anomalous. Although integrating Gemma into REACT increases its BLEU score by 51\%, it leads to a decline in the scores for the other two metrics. This is partly because the focus of evaluation differs among the three metrics. BLEU primarily measures the overlap between generated and reference texts, emphasizing precision. The other two metrics might assess aspects like relevance, fluency, or diversity, which could be adversely affected by the changes that improve BLEU scores. On the other hand, Gemma has the smallest parameter size among the three and has relatively weaker capabilities (according to the EvalPlus Leaderboard~\cite{evalplus}). The enhancement brought by REACT relies on the LLM's in-context learning ability, and Gemma's limited capabilities might prevent it from effectively following the instructions and also reliably benefiting from the guidance of the exemplar.

Overall, REACT still effectively enhances the performance of the other two LLMs across all three metrics. It even doubles Llama 3’s BLEU score, strongly demonstrating the effectiveness of REACT, proving that including an exemplar in the input for LLMs can guide better generation.

\begin{framed}
    \noindent \textbf{Key Findings of RQ3:} 
    \begin{itemize}
        \item Integrating LLMs into the REACT framework significantly enhances their CMG performance, especially in terms of BLEU scores.
        \item With REACT, Llama 3 achieves a 102\% improvement in BLEU score solely through prompting, without the need for training.
    \end{itemize}
\end{framed}

\subsection{RQ4: Ablation Study of the Hybrid Retriever}

This RQ intends to validate the effectiveness and necessity of the hybrid retriever within the REACT framework. We aim to determine whether the hybrid retriever can retrieve helpful exemplars. To achieve this, we conducted an ablation study by comparing the effects of different retrieval methods: random retrieval, single retrieval (using either BM25 or encoder), and hybrid retrieval. The generator for these experiments was the best-performing model identified in previous sections, CodeT5. By analyzing the results of these retrieval methods, we can assess whether the hybrid retriever truly works.

\begin{table}[t]
  \caption{Efficacy of different retrieval methods in REACT (Generator: CodeT5)}
  \label{tab:rq4}
  \begin{tabular}{lccc}
    \toprule
    Approach & BLEU & Rouge-L & METEOR \\
    \midrule
    No retrieval & 6.24 & 25.85 & 21.71\\
    Random & 6.05 & 26.05 & 21.62\\
    Only BM25 & 9.53 & 27.11 & 23.42 \\
    Only Encoder & 9.53 & 27.13 & 23.44 \\
    Hybrid & 9.68 & 27.20 & 23.65 \\
    \bottomrule
\end{tabular}
\end{table}
\subsubsection{Results.} Table~\ref{tab:rq4} presents the results of using different retrieval methods within the REACT framework, with CodeT5 as the generator. The table includes the metric scores for BLEU, Rouge-L, and METEOR. 
``No retrieval'' means not providing any exemplar, that is, direct application. ``Random retrieval'' means randomly selecting a diff-message pair from the source database to serve as an exemplar. The results indicate that using no retrieval achieves the lowest scores. Random retrieval does not aid the generation process and even slightly decreases performance compared to no retrieval. Both single retrieval methods (BM25 and encoder) significantly improve performance, achieving similar scores. However, the hybrid retrieval approach yields the highest scores across all metrics, indicating its superiority.

\subsubsection{Analysis.} The substantial improvement in performance when using the hybrid retriever demonstrates its effectiveness in retrieving similar exemplars. Specifically, the BLEU score increases by 60\% when comparing the hybrid retriever to random retrieval, the Rouge-L and METEOR scores also show noticeable improvements. The hybrid retriever effectively combines the strengths of both BM25 and encoder-based methods, retrieving more relevant diff-message pairs. This, in turn, provides better guidance for the generator, enabling it to produce more accurate commit messages. The results also reveal that while single retrieval methods are beneficial, the hybrid approach offers additional advantages, further justifying its inclusion in the REACT framework.

\begin{framed}
\noindent \textbf{Key Findings of RQ4:}
\begin{itemize}
\item The hybrid retriever can effectively retrieve helpful exemplars to guide generation, with a 60\% improvement in BLEU score compared to random retrieval.
\item Single retrieval methods (BM25 and encoder) provide substantial improvements, but the hybrid approach offers the highest performance across all metrics.
\end{itemize}
\end{framed}

\subsection{RQ5: Within Project Case Study}

\begin{figure}[t]
    \centering
    \includegraphics[width=\linewidth]{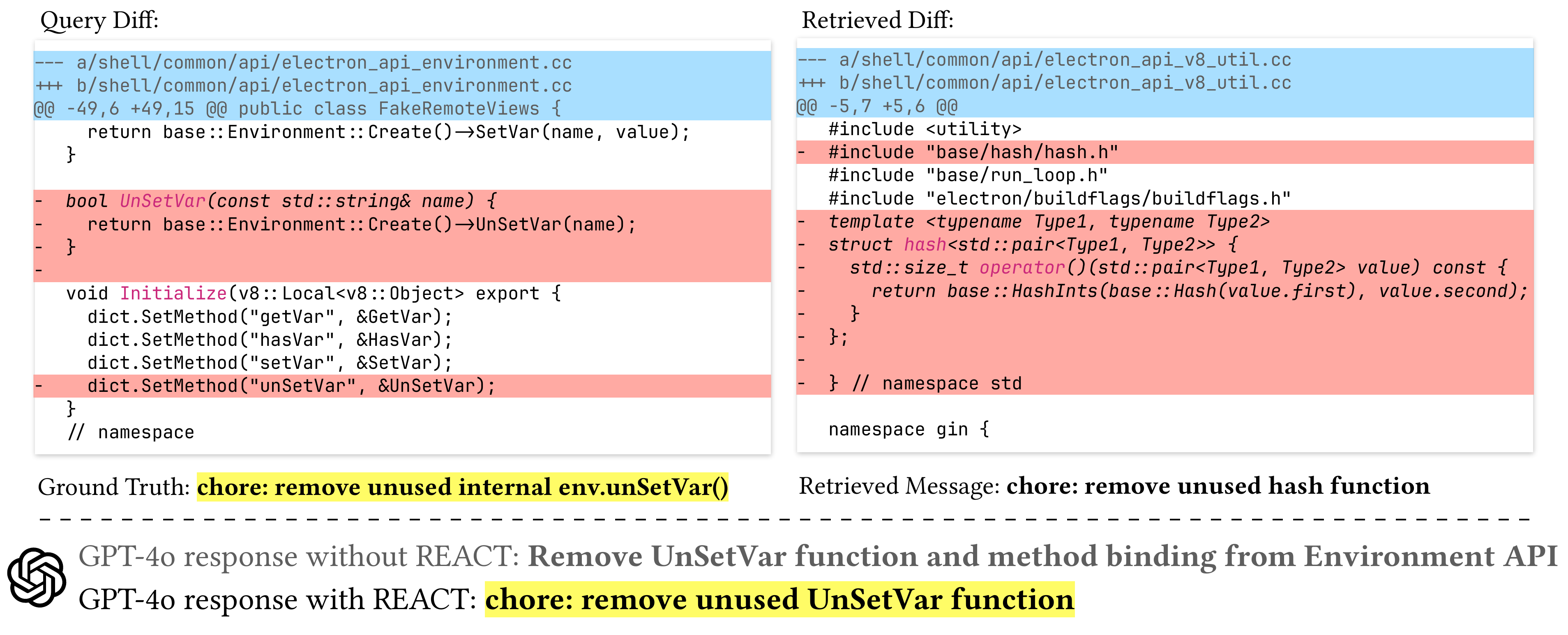}
    \caption{A case of with-in project study using GPT-4o integrated with REACT. Under the guidance of the retrieved exemplar, GPT-4o effectively adhered to the commit message writing conventions of the project, achieving a high degree of similarity with the ground truth.}
    \label{fig:within}
\end{figure}

The dataset used in this study was collected from multiple repositories. In fact, most of the existing widely-used CMG datasets are cross-project that do not differentiate between projects, meaning that the diff-message pairs retrieved in REACT may come from other projects. However, in general development practices, commit messages within a specific repository often follow similar conventions. When a developer writes a commit message for a project, they typically refer to the historical commit log to adopt a consistent style. In some cases, the repository’s code of conduct explicitly specifies the commit message conventions\footnote{Conventional Commits, \url{https://www.conventionalcommits.org/}} that should be followed. The REACT method proposed in this paper, a retrieval-augmented generation approach, intuitively proves to be more practical in this scenario since it can adapt the consistent writing style by following the guidance of retrieved exemplar. In this section, we selected an open-source project for a within-project case study, using experiments to further demonstrate the superiority of REACT in this scenario.

\subsubsection{Project Selection}

We selected an open-source project from GitHub as the case, \textbf{Electron}\footnote{\url{https://github.com/electron/electron}, visiting date: Sep 1, 2024.}, which meets the following criteria: it is highly popular, with 114k stars and 15.3k forks; it has a rich history of commits, exceeding 28,000 records; and it follows modern project organization and open-source contribution processes, with a comprehensive contribution guide and code of conduct. As a result, the quality of commit message writing is notably high.

\subsubsection{Dataset Collecting}

We scraped the historical commit records from the main branch as of January 1, 2020, and then filtered out commits with overly long code diffs and commit messages (>1000 tokens). Additionally, we excluded commits authored by bots, such as those by dependabot\footnote{\url{https://github.com/dependabot}}, which typically involve bumping a dependency package to a newer version. In the end, we obtained a dataset containing 3,604 diff-message pairs for the single project Electron.

\subsubsection{Results \& Analysis}

\begin{table}[h]
  \caption{Results of the with-in project experiments for Electron}
  \label{tab:within}
  \begin{tabular}{lccc}
    \toprule
    Approach & BLEU & Rouge-L & METEOR \\
    \midrule
    GPT4o & 3.41 & 23.79 & 13.03\\
    REACT(GPT4o) & \textbf{10.46} & \textbf{30.42} & \textbf{30.57}\\
    \bottomrule
\end{tabular}
\end{table}

We conducted a within-project experiment using GPT-4o. During the retrieval phase of the REACT method, we restricted the retrieval to similar examples only from the current project. As shown in Table~\ref{tab:within}, the experimental results indicate that in the within-project application scenario, the enhancement to the three metric scores provided by REACT is highly significant. It is able to generate commit messages that closely align with the project’s writing style, gaining higher metric scores than directly generating messages without REACT. Figure~\ref{fig:within} shows that GPT-4o, based on the retrieved exemplar, learned the commit message writing convention of this project as ``[action]: [description]''. The commit message generated by REACT successfully emulated this style. This further highlights the practical significance of REACT in real-world within-project scenarios, demonstrating that it not only significantly improves metric scores of CMG but also produces more usable commit messages that can be effortlessly adopted by developers.

\begin{framed}
\noindent \textbf{Key Findings of RQ5:}
\begin{itemize}
\item REACT enhances commit message generation in within-project scenarios by adapting to project-specific conventions through examples retrieved from the current project, making the messages more usable and aligned with project standards.
\end{itemize}
\end{framed}

\section{Implications}
\label{sec:6}
According to the findings of research questions and analysis of experimental results, we discuss the implications for the community:

\paragraph{\textbf{Enhanced Efficiency and Quality in CMG}} One of the most direct implications is the potential for enhanced efficiency and quality in CMG within software development processes. Writing commit messages is a crucial but often tedious task for developers. By automating this process with higher accuracy and relevance, CMG techniques can save developers considerable time and effort. Based on the results of this paper, our proposed approach achieved the highest metric scores compared to all baselines for the selected dataset. This achievement further elevates the technique of automatic CMG to a level of practical value.

\paragraph{\textbf{Broader Applicability of Retrieval-Augmented Generation (RAG)}} The experimental results indicate that REACT can broadly and significantly enhance CMG, further validating the effectiveness of the RAG paradigm across various tasks. In fact, RAG has already become one of the best practices for text generation tasks in multiple domains~\cite{rag}, proving to be a simple yet effective concept. Our work further demonstrates the effectiveness of RAG in the CMG task and provides insights for future research.

\paragraph{\textbf{Advantages and Superiority of CLMs}} The results of Section~\ref{sec:rq1} indicate that by applying CLMs directly to the CMG task through fine-tuning or prompting, they have already surpassed all existing baselines. This raises a question: Do we really need to train a specialized model from scratch? The demonstrated advantages and superiority of CLMs in the CMG task highlight their potential for other code-related tasks in software engineering. Researchers can investigate the integration of CLMs with various other software engineering tools, extending their utility beyond CMG.

\section{Threats to Validity}
\label{sec:7}

We discuss the potential threats to the validity of this study according to the guidelines proposed by~\cite{runeson2009guidelines}, and the impact that these threats may have on our study.

\paragraph{\textbf{Internal Validity.}} A primary threat to internal validity is the inherent randomness of LLMs. We keep all parameters, including temperature, at their default settings when using them. However, due to the LLMs’ inherent uncertainty, the generated results may vary with each run. To mitigate this threat, future work could involve repeating the experiments multiple times and averaging the results to reduce the impact of randomness. Despite this variability, the significant improvements observed in the experiments with LLMs in our study are sufficient to draw reliable conclusions.

\paragraph{\textbf{External Validity.}} A potential threat to validity arises from generalizability. Although our chosen dataset is widely used~\cite{cc2vec, codisum, fira, cmtgen, nngen,corec}, it only includes Java language and does not investigate the model's generalizability to other languages. We conducted our study on four PLMs and three LLMs with varying parameter sizes and generation capabilities, but we cannot confirm that the results generalize to other models. Future work could involve broader experiments to validate our approach.

\paragraph{\textbf{Construct Validity.}} A potential threat to construct validity in this research arises from the evaluation metrics used to assess the quality of the generated commit messages. The three metrics employed primarily measure text similarity between the generated messages and the reference messages. While these metrics are widely used in NLP tasks, they may not fully capture the essential qualities that constitute a high-quality commit message. Future work could consider incorporating additional evaluation methods, such as human evaluations.

\section{Conclusions}
\label{sec:8}

This paper proposed REACT, a retrieval-augmented framework designed to enhance commit message generation by effectively integrating advanced retrieval techniques with various language models with code capabilities. Through comprehensive evaluations, we demonstrated that REACT significantly improves the performance of both pre-trained language models and large language models on the commit message generation task. Experimental results indicate that when CodeT5 is integrated into REACT, its metric scores outperform all baselines and achieve the new SOTA. These findings underscore the efficacy of leveraging retrieval-augmented generation in conjunction with the rich prior knowledge embedded in CLMs, providing a robust and efficient solution for automated commit message generation.

For future work, we will consider incorporating additional information beyond retrieving similar exemplars to enhance the input. For instance, for issue-related commits, we can include the issue’s description and discussion information to aid in generating commit messages. Meanwhile, the RAG paradigm can be extended to more software engineering tasks.


\begin{acks}
This work has been partially supported by the National Natural Science Foundation of China (NSFC) with Grant Nos. 62032016 and 62172311.
\end{acks}

\bibliographystyle{ACM-Reference-Format}
\bibliography{sample-base}

\end{document}